\relax
\documentstyle [12pt]{article}
\begin {document}
\bibliographystyle {plain}

\title{\bf Semiclassical  Solution of One Dimensional
Model of  Kondo Insulator}
\author {A. M. Tsvelik}
\maketitle
\begin {verse}
$Department~ of~ Physics,~ University~ of~ Oxford,~ 1~ Keble~ Road,$
\\
$Oxford,~OX1~ 3NP,~ UK$\\
\end{verse}
\begin{abstract}
\par
The  model of Kondo chain with $M$-fold degenerate
band of conduction electrons of spin 1/2 interacting with localized
spins $S$ is studied for the case when the electronic band is
half filled. It is shown that
the spectrum of spin excitations in the continuous limit is
described by the O(3) nonlinear sigma model with the topological term
with $\theta = \pi(2S - M)$. For a case $|M - 2S| = $(even)
the system is an insulator and  single electron excitations
at low energies are massive spin polarons.  Otherwise the density of states
has a pseudogap and vanishes only at the Fermi level.
The relevance of this picture to
higher dimensional Kondo insulators is discussed.
\end{abstract}

PACS numbers: 74.65.+n, 75.10. Jm, 75.25.+z
\sloppy
\par
 The problem of co-existence of delocalized and localized electrons
in crystals  remains one of the biggest unsolved
problems in the condensed matter physics. The only part of this
problem which is well understood concerns a situation when localized
electrons are represented via a single local magnetic moment (the
Kondo problem). In this case the local moment is screened at low
temperatures by conduction electrons
and the ground state is a singlet. The singlet formation is a
non-perturbative process and the relevant energy scale (the Kondo
temperature) is exponentially small in the exchange
coupling constant. It is still unclear how
conduction and localized electrons reconcile with each other
when the local moments are arranged regularly
(Kondo lattice problem). Empirically Kondo lattices
resemble metals with very small Fermi energies of order of several
degrees.
It is widely believed that conduction and localized electrons
in Kondo lattices hybridize at low temperatures to create a single
narrow band. It is not at all clear, however, how this
hybridization develops. In particular, it is not clear whether
the localized electrons contribute to the volume of  Fermi sea.
If the answer is positive, a system with one conduction
electron and one spin per a unit cell must be an insulator.
The available experimental data apparently support this point
of view: all compounds with odd number of conduction  electrons
per spin are insulators$^1$. This class of compounds is now
known as Kondo insulators. At low temperatures they behave
as semiconductors with very small gaps of the order of several degrees.
The marked exception
is $FeSi$ where the value of  gap is estimated as $\sim 700K$$^2$.
The conservative approach to Kondo insulators would be to
calculate their band structure treating the on-site Coulomb
repulsion $U$
as a perturbation. The advantage of this approximation is that
one gets an insulating state already in the zeroth order in $U$.
The disadvantage is that it contradicts  the principles
of perturbation theory which prescibe that
the strongest interactions  are  taken into account first.
It turns out also that
the  pragmatic sacrifice of principles does not lead to a satisfactory
description of the experimental data: the band theory fails to
explain many experimental observations (see Ref. 2 for a discussion).

 In this letter  I study a one dimensional model of the Kondo lattice at the
half filling. I show that the insulating state forms not due to
a hybridization of conduction electrons with local moments,
but as a result
of strong antiferromagnetic fluctuations. Due to the local doubling
of period of the original lattice the conduction electrons become
heavier. An interaction of these heavy electrons with spin kinks
lead to the formation of massive spin polarons.
This scenario does not require a  global antiferromagnetic order,
just the contrary -
the spin ground state remains disordered with
a finite correlation length. I
suggest that such scenario can be generalized for higher dimensions.
Kondo insulators in this case are either antiferromagnets (then they
have a true gap), or spin fluids with a strongly
enhanced staggered susceptibility. In the latter case instead of
a real gap there is a pseudogap - a drop in the density of states
on the Fermi level. The
recent numerical calculations of Yu et. al.$^3$ also demonstrate
a sharp enhancement of the staggered susceptibility in one-dimensional
Kondo insulators.


 As a model of one-dimensional Kondo insulator I
consider the model of Kondo chain at half filling governed by the
following Hamiltonian:
\begin{eqnarray}
H =
\sum_r \sum_{\alpha = 1}^M
[ - \frac{1}{2}(c^+_{r + 1,\alpha, a} c_{r,\alpha, a}
+ c^+_{r,\alpha, a} c_{r + 1,\alpha, a}) +
J (c^+_{r,\alpha, a}\hat{\vec\sigma}_{ab}c_{r,\alpha, b})
\vec S_r] \label{eq:model}
\end{eqnarray}
It describes an $M$-fold degenerate band of electrons with spin S = 1/2
interacting with local spins S.
In what follows I shall use the path integral formalism. The path
integral representation  for spins has been discussed in details
by many authors. I would refer a reader to the book of Fradkin$^4$.
In the path integral  spins are treated as  classical variables
$\vec S = S\vec m$ ($\vec m^2 = 1$); the corresponding Euclidean
action for the model (1) is given by:
\begin{eqnarray}
A = \nonumber\\
\int d\tau \{\sum_r[iS\int_0^1 du(\vec m_r(u,\tau)[\partial_u\vec
m_r(u,\tau)\times\partial_{\tau}\vec m_r(u,\tau)]) + c^*_{r,\alpha,
a}\partial_{\tau}c_{r,\alpha, a}] - \nonumber\\
H(c^*,c; S\vec m)\} \label{eq:action}
\end{eqnarray}
The first term is the  spin Berry phase responsible for  the correct
quantization
of local spins. Since the integrand in the
Berry phase is a total derivative, the
integral depends only on the value of $\vec m$ on the boundary, i.e.
on $\vec m(u = 0, \tau) = \vec m(\tau), \vec m(u = 1,\tau)
= (1,0,0)$. The introduction of the additional variable $u$ is a
price one has to pay for the fact that the
Berry phase cannot be written as a local
functional of $\vec m_r(\tau)$.

 I shall follow the semiclassical approach assuming  that all fields
can be separated into fast and slow components. The fast components
then will be  integrated out and as a result
I shall obtain an  effective action for the
slow ones. This approach is self-consistent if the obtained
correlation length for spins
is  much larger then the lattice spacing. In ordinary antiferromagnets
this requirement is fulfilled only for large spins $S >> 1$.
As we shall see later, the Kondo chain is semiclassical
even for $S = 1/2$ provided the exchange integral is small $JM << 1$.
 I suggest the following decomposition of
variables:
\begin{eqnarray}
\vec m_r = a\vec k(x) + (-1)^r\vec n(x)\sqrt{1 - a^2\vec k(x)^2}, \nonumber\\
(\vec k \vec n) = 0 , \nonumber\\
c_r = i^r\psi_R(x) + (-i)^r\psi_L(x) \label{eq:separ}
\end{eqnarray}
where $|\vec k|a << 1$ is the fastly varying ferromagnetic component of the
local magnetization. Substituting Eqs.(~\ref{eq:separ})
into Eq.(~\ref{eq:action}) and keeping only  non-oscillatory
terms, I get:
\begin{eqnarray}
A = \int d\tau dx L, \nonumber\\
L = iS(\vec k[\vec n\times\partial_{\tau}\vec n]) +  \bar\psi_j
[i\gamma_{\mu}\partial_{\mu}\hat I + JS(\hat{\vec\sigma}\vec n(x))
\sqrt{1 - a^2\vec k(x)^2}]\psi_j + \nonumber\\
2\pi S\times(top-term) \label{eq:action1}
\end{eqnarray}
where
\begin{equation}
Top-term = \frac{i}{8\pi}\int d\tau dx \epsilon_{\mu\nu}\left(\vec
n[\partial_{\mu}\vec n\times\partial_{\nu}\vec n]\right) \label{eq:top}
\end{equation}
is the topological term first derived by Haldane$^5$. As far as the
non-electronic part of the action is concerned, my
derivation repeats the one presented in Ref. 4. The interaction of
electrons with ferromagnetic fluctuations has been omitted; it can be
shown that at small $JM << 1$ it gives insignificant corrections.

The fermionic determinant is calculated later (see
Eq.(~\ref{eq:deter})).
Besides of the trivial static part it
contains the topological term, but with $ - M$ instead of $2S$. This
is what one should expect: this change reflects the fact that
local spins couple with conduction electrons to give the total spin
$S - M/2$.
Substituting the expression (~\ref{eq:det1}) into
Eq.(~\ref{eq:action1}), I get:
\begin{eqnarray}
L = iS(\vec k[\vec n\times\partial_{\tau}\vec n]) +  \frac{M}{2\pi}[
(\partial_{x}\vec n)^2 + (\partial_{\tau}\vec n)^2] + \nonumber\\
\frac{2M}{\pi}(JS)^2\ln\frac{1}{JS}(\vec k)^2 +
\pi(2S - M)\times(top-term)
\end{eqnarray}
Integrating over fast ferromagnetic fluctuations described by $\vec k$,
I get
\begin{eqnarray}
A = \frac{M}{2\pi}\int d\tau dx\left[v^{-2}(\partial_{\tau}\vec n)^2 +
(\partial_x\vec n)^2\right] +
\pi(2S - M)\times(top-term) \label{eq:O(3)}\\
v^{-2} =  1 + \frac{2\pi^2}{J^2M^2\ln(1/JS)}
\end{eqnarray}
After the rescaling of the coordinates $v\tau = x_0, x = x_1$ I get
the action of the O(3) nonlinear sigma model with the
dimensionless coupling constant
\begin{equation}
g = \frac{\pi v}{M} = \frac{\pi}{\sqrt{M^2 + \frac{2\pi^2}{J^2\ln(1/JS)}}}
\end{equation}
This constant is small at $JM << 1$ which justifies the entire
semiclassical approach.
At $|M - 2S| =$ (even) one can omit the topological term. In this
case  the model Eq.(~\ref{eq:O(3)}) is the ordinary O(3) nonlinear sigma
model. This model has
a disordered ground state with the spectral gap$^{6,7}$
\begin{equation}
\Delta = Jg^{-1}\exp[- 2\pi/g] \label{eq:gap}
\end{equation}
and the correlation length $\xi \sim Ja/\Delta >> a$. If $|M - 2S|$ = (odd) the
topological term is essential$^5$. The model
becomes critical and the correlation functions of staggered magnetization
have a power law decay. The specific heat is linear at small temperatures
without requiring, however, the single electron density of
states to be  constant at the Fermi level.


 Now I shall evaluate the fermionic determinant
\begin{eqnarray}
D[g] = MTr\ln[i\gamma_{\mu}\partial_{\mu} +
(1 + i\gamma_5)mg/2 + (1 - i\gamma_5)mg^+/2] \label{eq:det}
\end{eqnarray}
where $g$ is a matrix from $SU(N)$ group and $m$ is
some constant energy scale
(in the context of the model (~\ref{eq:model}) $m = JS$ and $g =
(\vec\sigma\vec n)$).
N = 2 in the original problem, but it is
worth to do the calculation for general N. I shall study  the
expansion of the determinant (~\ref{eq:det})
in terms of  $m^{-1}\nabla g$. The first terms of this expansion are
independent of $m$ and survive even at $m \rightarrow \infty$.
I claim that the gradient  expansion  contains
a Berry phase. To prove this point I take a route which may seem exotic, but
I do not know any better way to get the right answer. As a preliminary step
I consider the chiral Gross-Neveu model with
the $U(M)\times SU(N)$-symmetry described by the following  action:
\begin{eqnarray}
A = \nonumber\\
\int d^2x \{i\bar\eta_{a,\alpha}\gamma_{\mu}\partial_{\mu}\eta_{a,\alpha}  -
\frac{c}{2}[(\bar\eta_{a,\alpha}\eta_{b,\alpha})(\bar\eta_{b,\beta}\eta_{a,\beta}) -
(\bar\eta_{a,\alpha}\gamma_5\eta_{b,\alpha})(\bar\eta_{b,\beta}\gamma_5
\eta_{a,\beta})]\label{eq:nonabel}
\end{eqnarray}
The Greek indices belong to the group $SU(M)$ and the Latin ones to $SU(N)$.
To avoid a confusion I emphasise that this model {\it is not}
equivalent to the original model and I consider it only because
the  effective  action for its low energy excitations are given
by the determinant (~\ref{eq:det}). In order to show that
I introduce the auxilary field
 $Q_{ab}$ and  decouple the interaction
term by the Hubbard-Stratonovich transformation. Interacting formally over
the fermions we obtain the partition function for the tensor $Q_{ab}$:
\begin{eqnarray}
Z = \int DQ^+DQ\exp( - \int d^2x L),\nonumber\\
L =  \frac{1}{2c}TrQ^+Q  - \nonumber\\
MTr\ln[i\gamma_{\mu}\partial_{\mu} +
(1 + i\gamma_5)Q/2 + (1 - i\gamma_5)Q^+/2] \label{eq:z}
\end{eqnarray}
As in the standard $U(N)$-invariant Gross-Neveu model,
the effective action has a saddle point with
respect to $Q^+Q$ and  fluctuations of  $\det Q$ are massive.
This can be shown in the standard fashion. Assuming
 that  the saddle point configuration of $Q$ is coordinate independent and
as such can be chosen as a diagonal real matrix: $Q(x) = diag(\lambda_1, ...
\lambda_r)$, I calculate  the
density of effective action (~\ref{eq:z}) on this configuration:
\begin{equation}
A_{eff} = \sum_{a}[\lambda_{a}^2/2c +
\frac{M\lambda^2_{a}}{2\pi}\ln\frac{|\lambda_{a}|}{\Lambda}]
\end{equation}
where $\Lambda$ is the ultraviolet cut-off.
The saddle point value of $Q$, being a point of minimum of this function,
satisfies the following equation:
\begin{equation}
\lambda_{a}/c + \frac{M\lambda_{a}}{\pi}
\ln\frac{|\lambda_{a}|}{\Lambda} = 0
\end{equation}
which solution is
\begin{equation}
\lambda_{a} = \Lambda\exp[- \pi/Mc] \equiv m \label{eq:mass}
\end{equation}
The performed calculation suggests that for slowly varying fields $Q$
one can substitute $Q_{ab}$ in the $Tr\ln$ in Eq. (~\ref{eq:z})
by $mg_{ab}$ where $g$ is an $SU(N)$ matrix. In other words, the effective
action for  excitations of the model (~\ref{eq:nonabel})
with energies $ << m$ coinsides with the fermionic determinant (~\ref{eq:det})
with $m$ given by Eq.(~\ref{eq:mass}).

 On the next step of the derivation I use the fact that due to the
identity $2\tau_1^a\tau_2^a = 1/2 - P_{12}$, where $P_{12}$ is the
permutation operator, the
chiral Gross-Neveu model (~\ref{eq:nonabel}) can be rewritten as
the model with
current-current interaction:
\begin{eqnarray}
A = \int d^2x \{i\bar\eta_{a,\alpha}\gamma_{\mu}\partial_{\mu}\eta_{a,\alpha}
+ 4c J_{\mu}^{\lambda}J^{\mu {\lambda}}\}, \nonumber\\
 J_{\mu}^{\lambda} =
(\bar\eta_{a,\alpha}\gamma_{\mu}\tau_{\alpha\beta}^{\lambda}\eta_{a,\beta})
\label{eq:nonabel1}
\end{eqnarray}
where $\tau^r$ are matrices - generators of the $SU(M)$ group.
The two models differ by a term containing a diagonal scattering. This term
does not renormalize and therefore is not important. Now I apply to
the model (~\ref{eq:nonabel1}) the non-Abelian bosonization procedure
suggested by Witten$^8$ (see also the book $^9$).
Namely, I rewrite
its Hamiltonian in the Sugawara form:
\begin{eqnarray}
H  = H_{U(1)} + H_{SU(N)} + H_{SU(M)}, \label{eq:ham1}\\
H_{U(1)} = \pi\int dx[:J_R(x)J_R(x): + :J_L(x)J_L(x):]\\
 H_{SU(N)} = \frac{2\pi}{(N + M)}\sum_{i = 1}^{G_N}\int dx [:J^i_R(x)J^i_R(x):
+ :J^i_L(x)J^i_L(x):] \\
 H_{SU(M)} =  \nonumber\\
\sum_{{\lambda} = 1}^{G_M}\int dx[\frac{2\pi}{(N +
M)}(:J^{\lambda}_R(x)J^{\lambda}_R(x): + :J^{\lambda}_L(x)J^{\lambda}_L(x):) +
4c :J^{\lambda}_R(x)J^{\lambda}_L(x):]
\end{eqnarray}
where I have introduced the chiral currents satisfying the
Kac-Moody algebra; for the group $SU(N)$ the
corresponding defenition is
\begin{eqnarray}
J_R^i = \eta_{R,a\alpha}^+t^i_{ab}\eta_{R,b\alpha},\nonumber\\
J_L^i = \eta_{L,a\alpha}^+t^i_{ab}\eta_{L,b\alpha}
\label{eq:currents}
\end{eqnarray}
where $t^i$ are the generators of the  $SU(N)$ (spin)
group and $\eta_R, \eta_L$
are right and left components of the Dirac spinor $\eta$.  $J$ and
$J^{\lambda}$
are the $U(1)$ and the $SU(M)$ (flavour)
currents defined as in Eq.(~\ref{eq:currents}), but with  generators of
the corresponding algebras. $G_N = N^2 - 1$ and $G_M = M^2 - 1$ are
the total number of generators of the $su(N)$ and the $su(M)$ algebras.
Currents from different algebras commute. Therefore the Hamiltonian
(~\ref{eq:ham1})
is a sum of three mutually commuting operators. Now notice that
the interaction term containing right and left $SU(M)$ currents
do not affect the
spectra of $SU(M)$-singlets. It is well known that the spectrum of
$H_{SU(M)}$ is gapful for the given sign of the coupling constant
(see, for example, Ref. 10). From the previous discussion we
can conclude that  the gapful excitations correspond to fluctuations
of $\det Q$.
Therefore
the spectrum below the gap $m$ is described by the rest of
the Hamiltonian (~\ref{eq:ham1}), in other words by
\begin{eqnarray}
H_{eff} =  H_{U(1)} + H_{SU(N)}; \label{eq:u(n)}\\
H_{U(1)} =  \pi\int dx[ :J_R(x)J_R(x): + :J_L(x)J_L(x):]; \\
H_{SU(N)} = \int dx \frac{2\pi}{(N + M)}\sum_{a = 1}^{G_N}( :J^i_R(x)J^i_R(x):
+ :J^i_L(x)J^i_L(x):) \label{eq:wzwh}
\end{eqnarray}
The Hamiltonian (~\ref{eq:wzwh}) is the Hamiltonian of the Wess-Zumino-Witten
model on the group $SU(N)$.
Its spectrum is the subsector of the free fermionic spectrum generated
by the $SU(N)$ current operators.  The model is conformally invariant and
exactly solvable$^{11,12}$. In order to extract from these results
an expression for the determinant (~\ref{eq:det}), I  rewrite the model
(~\ref{eq:u(n)}) in the Lagrange representation$^{8-12}$:
\begin{equation}
A_{U(1)} = \frac{NM}{8\pi}\int d^2x (\partial_{\mu}\phi)^2 \label{eq:bosons}
\end{equation}
\begin{equation}
A_{SU(N)} = \int d^2x \{\frac{M}{16\pi}Tr(\partial_{\mu}g^+\partial_{\mu}g) +
\frac{M}{24\pi}
\int_0^1 d\xi \epsilon_{abc}Tr(g^+\partial_a gg^+\partial_b gg^+\partial_c g)\}
\label{eq:wzw}
\end{equation}
where $g$ is a matrix from the $SU(N)$ group ($g^+g = I, \det g = 1$).
The second term in the right hand side of
Eq.(~\ref{eq:wzw}) is called the Wess-Zumino term. This term is topological.
Despite of the fact that it is written as an integral including the
additional dimension, its actual value (modulus $2\pi iN$) depends on
the boundary values of $g(x,\xi = 0) = g(x)$ ($g(x,\xi = 1) = 0$). This
property follows from the fact that the Wess-Zumino term is proportional to
the
integral of the
Jacobian of transformation
from the three dimensional euclidian coordinates to the group coordinates,
and so it is  a total derivative in disguise.

 Now we can recognize in Eq.(~\ref{eq:wzw}). Indeed, its first part (except of
the Wess-Zumino term)
represents the first term in the gradient expansion of $Tr\ln$-term in
Eq.(~\ref{eq:z}). Indeed,
at small momenta $|p| << m$ this term is equal to
\begin{equation}
\frac{M}{16\pi m^2}\int d^2x Tr(\partial_{\mu}Q^+\partial_{\mu}Q)
\label{eq:naive}
\end{equation}
Now let us write
down the $Q$-field as follows:
\[
Q(x) = mg(x)e^{i\sqrt M\phi(x)},
\]
where $g$ belongs to the $SU(N)$ group. Substituting this expression
into Eq.(~\ref{eq:naive}) and
taking into account that $Trg^+\partial_{\mu}g = 0$ as a trace of an element of
the
algebra,  we
reproduce Eq.(~\ref{eq:bosons}) and the first term of Eq.(~\ref{eq:wzw}).
It comes
not entirely  unexpected that the naive gradient expansion misses the important
Wess-Zumino term.
Being a Berry phase this term requires special care and
cannot be derived from a gradient expansion. In order to avoid possible
calculational difficulties I have resorted to the exact solution
which gives us the following interesting expression for the determinant
(~\ref{eq:det}):
\begin{eqnarray}
D[g] = A_{SU(N)}[g] +  \frac{Mm^2}{2\pi}\ln(\Lambda/m) + O(m^{-2})
\label{eq:deter}
\end{eqnarray}
where $A_{SU(N)}[g]$ is given by Eq.(~\ref{eq:wzw}) and the second term
represents the static part of the determinant.
For the particular case $g = (\vec{\hat\sigma}
\vec n)$ I get
\begin{equation}
A_{SU(2)}[(\vec{\hat\sigma}\vec n)] = \frac{1}{2\pi}(\partial_{\mu}\vec n)^2 +
\pi(Top-
term) \label{eq:det1}
\end{equation}


 Let us discuss the excitation spectrum. I shall do it only
for $M = 1$. In this case the original
model has a combined symmetry $SU(2)\times SU(2)$ (the
additional $SU(2)$-symmetry arises as a particle-hole symmetry at
the half-filling), excitations carry two quantum numbers - spin S,
and an isotopic spin $I$. We have established
that spin excitations, i.e. excitations with $I = 0$,
are described by the nonlinear
sigma model with the topological term(~\ref{eq:O(3)}). The leading
contributions to the low energy dynamics come from antiferromagnetic
fluctuations which agrees with the results of Ref. 3. The
corresponding energy scale (~\ref{eq:gap}) is formally resembles
the expression for the Kondo temperature. $m(J)$ is larger, however,
due to the presence of a large logarithm. Therefore the
RKKY interaction plays a  stronger role than the Kondo screening -
it also agrees with the conclusions of Ref. 3. The
topological term can be omitted if $|M - 2S|$ = (even). In particular,
it cancels for the most physical case $S = 1/2, M = 1$. The low
lying magnetic excitations
are in this case  massive triplets, as it is for the O(3) nonlinear
sigma model$^{6,7}$. This picture is in a
qualitative agreement with the strong coupling limit of the model
(~\ref{eq:model}). Indeed, for $J >> 1$ the ground state of the Kondo
chain consists of local singlets. Excited states are triplets and are
separated by the gap $\sim J$ from the ground state.

 In order  to describe the fermionic excitations which have S = 1/2,$I
= 1/2$. It follows from (~\ref{eq:action1}) that
the fermionic fields live in a slowly fluctuating field
$JS\{\vec{\hat\sigma}\vec n(x,\tau)\}$. For constant  $\vec n$
the electrons would have
a spectral gap $JS$.
There are states in the gap, however, and they are bound
states of electrons and solitons of the unit vector field
$\vec n$. Such situation is typical for relativistic
field theories and for  the $SU(N)$
chiral Gross-Neveu model it was discussed  in Ref. 13.
Let us consider a slow static configuration $\vec n(x)$ such
that $\vec n(- \infty)$ is antiparallel to the $z$-axis and
$\vec n(\infty)$ is parallel to it. Then a straightforward
calculation shows that there is an electronic bound state on this domain
wall with the zero energy. The corresponding wave function is equal
to
\begin{equation}
\Psi(x) = \hat T\exp[ - \int_0^{\infty} dy(\vec\sigma\vec
n(y))]\Psi(0)
\end{equation}
The total energy of this bound state is equal to the energy necessary
to create the domain wall, which is the gap for the spin excitations.
 Therefore in the energy
interval between $JS$ and $m$  single electron
excitations are massive spin polarons. It is no longer the case if
$|M - 2S|$ = (odd). The nonlinear sigma model becomes critical and
supposedly belongs to the universality class of the isotropic S = 1/2
Heisenberg chain. In the critical phase the
spin solitons do not have a fixed scale and the corresponding bound
states can have an arbitrary small energy. It is reasonable to
suggest that the single particle density of states in this case
has a pseudogap on the Fermi surface decaying as a power law:
$\rho(\omega) \sim |\omega|^{\alpha}$.

 We can generalize this one dimensional picture for higher dimensions. Suppose
we have a Kondo lattice in three dimensions. Then at half filling we can
achieve the insulating state by the periodicity doubling via
an antiferromagnetic phase transition. Suppose, however, that this transition
does not
occur, but the system is very close to it. Then the low lying excitations
are spin polarons. Whether they have a gap or not depends on
the state of the spin system. For the one dimensional O(3) nonlinear
sigma model the static magnetic susceptibility
is strongly  enchanced at the antiferromagnetic wave
vector $q = \pi$ ($\chi(\pi) \sim m^{-1}$). In the same time it is
zero at $q = 0$. The corresponding measurements in $Ce_3Bi_4Pt_3$
show a pronounced drop in $\chi(0)$ at low temperatures and the
neutron measurements show a spectral gap$^{1}$. It indicates that
the ground state in Kondo insulators is magnetically disordered.
I postpone a discussion   of three dimensional
problem to further publications.


 This work was started during my visit to
the Department of Physics of Chalmers University in Goteborg (Sweden). I am
grateful to Prof. S. Ostlund and the condensed matter group in
Chalmers for the kind hospitality. I am grateful to A.
Nersesyan and G. Japaridze for inspirational conversations, for P.
Coleman for the valuable criticism, to Derek Lee for reading the
manuscript  and to C. Yu for sending me her
preprint.


\end{document}